\documentclass[]{spie}  

\usepackage{amsmath,amsfonts,amssymb}
\usepackage{graphicx}
\usepackage{upgreek}
\usepackage{xspace}
\usepackage{marvosym}
\usepackage[colorlinks=true, allcolors=blue]{hyperref}

\newcommand{\umux}{$\upmu$mux\xspace}

\title{A simulation suite for readout with SMuRF tone-tracking electronics}

\author[a]{Cyndia Yu}
\author[b]{Zeeshan Ahmed}
\author[b]{J. Mitch D'Ewart}
\author[b]{Josef C. Frisch}
\author[b]{Shawn W. Henderson}
\author[c]{Max Silva-Feaver}
\affil[a]{Stanford University Department of Physics, Stanford, CA 94305}
\affil[b]{SLAC National Accelerator Laboratory, Menlo Park, CA 94025}
\affil[c]{University of San Diego Department of Physics, La Jolla, CA 92093}

\authorinfo{\Letter~\texttt{cyndiayu@stanford.edu}}

\begin{document} 
\maketitle

\begin{abstract}
We present the details of a simulation suite for modeling the effects of readout with SLAC Microresonator RF (SMuRF) electronics. 
The SMuRF electronics are a warm readout and control system for use with superconducting microwave resonator-based detector systems. 
The system has been used with the BICEP/Keck program and will be used on the upcoming Simons Observatory and BICEP Array experiments. 
This simulation suite is a software implementation of the main SMuRF algorithms for offline analysis, modeling, and study. 
The firmware-implemented algorithms for calibration, resonator frequency estimation, and tone tracking present sources of potential bias or errors if not modeled properly. 
The simulator takes as input true detector signal, realistic resonator properties, and SMuRF-related user-controlled readout settings. 
It returns the final flux ramp-demodulated output of a detector timestream as would be passed to the experiment data acquisition system, enabling the analysis of the impact of readout-related parameters on the final science data. 
It is publicly available in Python with accompanying Jupyter notebooks for user tutorials. 
\end{abstract}


\section{Introduction}
\label{sec:intro}

Superconducting detectors are increasingly popular for a broad variety of applications due to their exquisite sensitivity, high resolution, and low noise.\cite{irwinhilton,enss05}
As these experiments seek to deploy ever-larger arrays to achieve their science goals, new multiplexing techniques are required to reduce thermal loading and integration complexity. 
Superconducting resonator-based detector or readout schemes transduce the detector signal into modulation of the amplitude and/or phase of a microwave resonator. 
The wide available readout bandwidth and high quality factors routinely achieved in nanofabrication allow for high channel count and/or high bandwidth readout of superconducting detectors, qubits, and other cryogenic devices. 
These systems require sophisticated RF readout electronics for interrogating the resonators and performing the digital signal processing necessary to infer the detector signal. 

SMuRF electronics are designed to meet the needs of large-format arrays read out with microwave SQUID multiplexing (\umux).\cite{irwin04,mates08}
The SMuRF system provides RF tone generation and readback, flux ramp signal generation and demodulation, and other subsystems such as DC signal generation and conditioning, data streaming, and timing synchronization that comprise a full experiment readout. 
The system has been demonstrated in an engineering run on the Keck Array and will be deployed on the upcoming Simons Observatory experiment.\cite{cukierman20,mccarrick21}
Laboratory-scale deployments are actively in use around the world for testing and characterization of \umux and KID systems. 

The SMuRF electronics are a highly complex and integrated system with hardware, firmware, and software components that all contribute to the overall transfer function and performance of the readout. 
In particular, the frequency error estimation, tone-tracking, and flux ramp demodulation functions rely on proper calibration, well-behaved resonators, and well-tuned user input parameters to arrive at faithful reconstructions of the detector signal.  
In these proceedings we present a simulation suite to enable the exploration and modeling of these effects. 
The code implements the per-channel firmware algorithms as they are used in the SMuRF electronics, enabling users to predict and study the effects of the SMuRF signal processing on science data. 
We note that the current software suite does not include multichannel effects, such as implementation of the polyphase filter banks and crosstalk. 
It also does not yet include any resonator or SQUID physics, such as response to varying input power levels, changing resonance shapes, or resonator/SQUID coupling effects, though these updates are planned for future releases. 

The format of these proceedings is as follows. 
In Sec.~\ref{sec:res} we outline the basics of superconducting resonator readout and microwave SQUID multiplexing as relevant for the simulation suite. 
We give a brief outline of the SMuRF electronics in Sec.~\ref{sec:smurf} with special focus on the digital signal processing algorithms implemented in the software suite. 
An overview of the software implementation is given in Sec.~\ref{sec:software}. 
Three example use cases enabled by this software suite are shown in Sec.~\ref{sec:ex}. 
We conclude with a discussion of future directions in Sec.~\ref{sec:conclusion}. 

\section{Introduction to Resonator Readout}
\label{sec:res}

The SMuRF electronics were broadly designed for use with superconducting microwave-frequency resonators, which are a popular choice for detector and/or readout element for a variety of applications.\cite{zmuidzinas12}
In particular, the system was optimized for use with the microwave SQUID multiplexer, a technology for reading out many cryogenic detectors with a small number of wires. 
However, it is broadly applicable to and has been used with other resonator technologies, most notably kinetic inductance detectors (KIDs). 
In this section we present some basic elements of microwave SQUID multiplexing and resonator readout that will be relevant in later sections. 

\subsection{The Microwave SQUID Multiplexer}
\label{subsec:umux}
\begin{figure}
    \centering
    \includegraphics[width=0.95\textwidth]{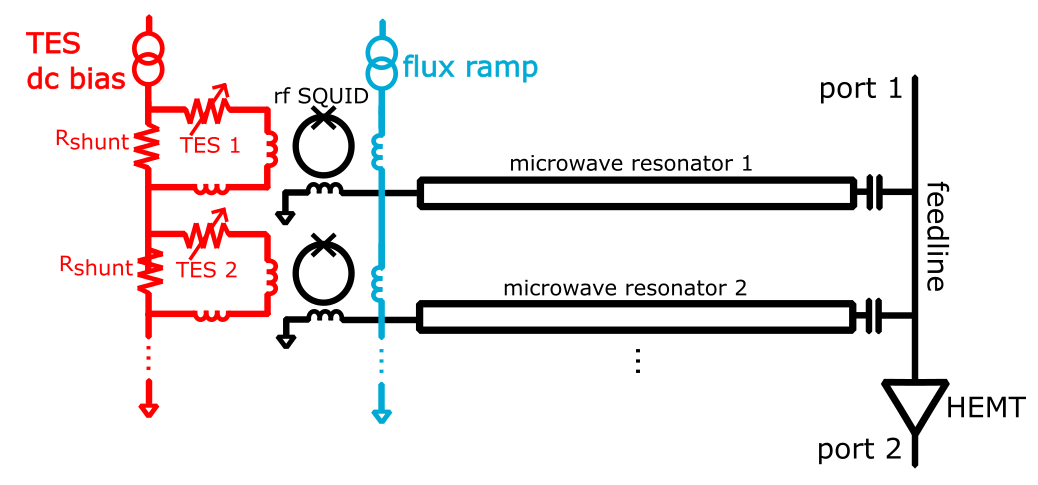}
    \caption{A schematic of the microwave SQUID multiplexer, shown here with two transition-edge sensor (TES) detectors and two resonators. In this scheme, a dc-biased TES (red) is uniquely coupled to a GHz-frequency resonator (black) via an rf SQUID. The detector signal modulates flux in the SQUID, which presents as a variable inductor coupled to the resonator. A common flux ramp line (blue) linearizes the SQUID response without the need for additional feedback lines.}
    \label{fig:umux}
\end{figure}

The microwave SQUID multiplexer consists of a superconducting GHz-frequency resonator coupled to a unique rf SQUID, which is in turn coupled to a unique cryogenic detector, typically a transition-edge sensor (TES) or a magnetic microcalorimeter (MMC).
A schematic is given in Figure~\ref{fig:umux}. 
A detector signal is thus transduced into a flux in the SQUID, which in turn modulates the effective inductance of the resonator and therefore its resonance frequency. 
Depending on available readout bandwidth and resonator quality factors, thousands of resonators may be coupled to a common feedline and read out simultaneously. 

To linearize the SQUID response without adding a unique feedback line for every detector channel, a common flux ramp line is coupled to all rf SQUIDs and modulated much faster than the anticipated detector signal.\cite{mates12}
This modulation shape is traditionally implemented as a sawtooth waveform with sufficient amplitude to sweep several flux quanta in the SQUID per ramp. 
Thus, the detector signal is transduced to a phase modulation of the flux ramp-modulated resonance frequency response. 

The SQUID response to input flux is typically described in terms of the resonance frequency deviation from some initial state. 
The shape of this response is quasi-sinusoidal, with the deviation from a pure sinusoid parametrized by $\lambda$ corresponding to the SQUID hysteresis parameter.\cite{mates11} 
The amplitude of the frequency swing $f_\text{pp}$ is set by the coupling factors between the SQUID and the resonator. 
To maximize SQUID responsivity without risking resonator bifurcation or pathological SQUID curves, \umux devices are traditionally designed such that the peak to peak frequency swing $f_\text{pp}$ is matched to the resonance bandwidth $\Delta f$. 
However, specific applications or variations in fabrication may yield more undercoupled or overcoupled devices, where the coupling sets $f_\text{pp}$. 
For \umux, we operate in the regime where resonance shapes do not change substantially with SQUID flux. 
This is in contrast to KIDs, which often change in both resonance frequency and quality factor with detector loading.\cite{zmuidzinas12}

Since \umux allows for the benefits of microwave resonator-based readout systems while maintaining separately designed and optimized detector arrays, it is an increasingly popular readout scheme for applications in search of larger detector counts and/or high bandwidth cryogenic detector readout. 
The technology has been used for astronomical applications with the MUSTANG-II instrument on the Green Bank Telescope and the Keck Array cosmic microwave background (CMB) experiment, and is planned or considered  for use on the upcoming Simons Observatory and Ali-CPT CMB experiments as well as future far-IR and x-ray flagship space missions.\cite{stanchfield16,cukierman20,galitzki18,salatino20,bennett19,nagler21}
Beyond astronomical applications, \umux has been demonstrated in gamma ray spectroscopy and is being explored for use with x-ray spectroscopy with light sources.\cite{sledgehammer}

\subsection{Resonator Readout}
\label{subsec:resreadout}
For a series of resonators coupled to a common transmission line, the state of the resonators is probed by exciting the resonators and measuring the response. 
In this section and throughout the text we refer to the measured resonator response as its forward transmission $S_{21}$; other schemes are easily mapped onto this case. 
While some schemes utilize a ringdown method for interrogating the resonator state, the associated nonlinearity and aliasing penalties drive many readout schemes to favor constantly monitoring the transmitted response of a comb of probe tones tuned to each resonance. 
Thus, in the discussion that follows we assume the resonance is excited via a spectrally pure probe tone. 

\begin{figure}
    \centering
    \includegraphics[width=0.95\textwidth]{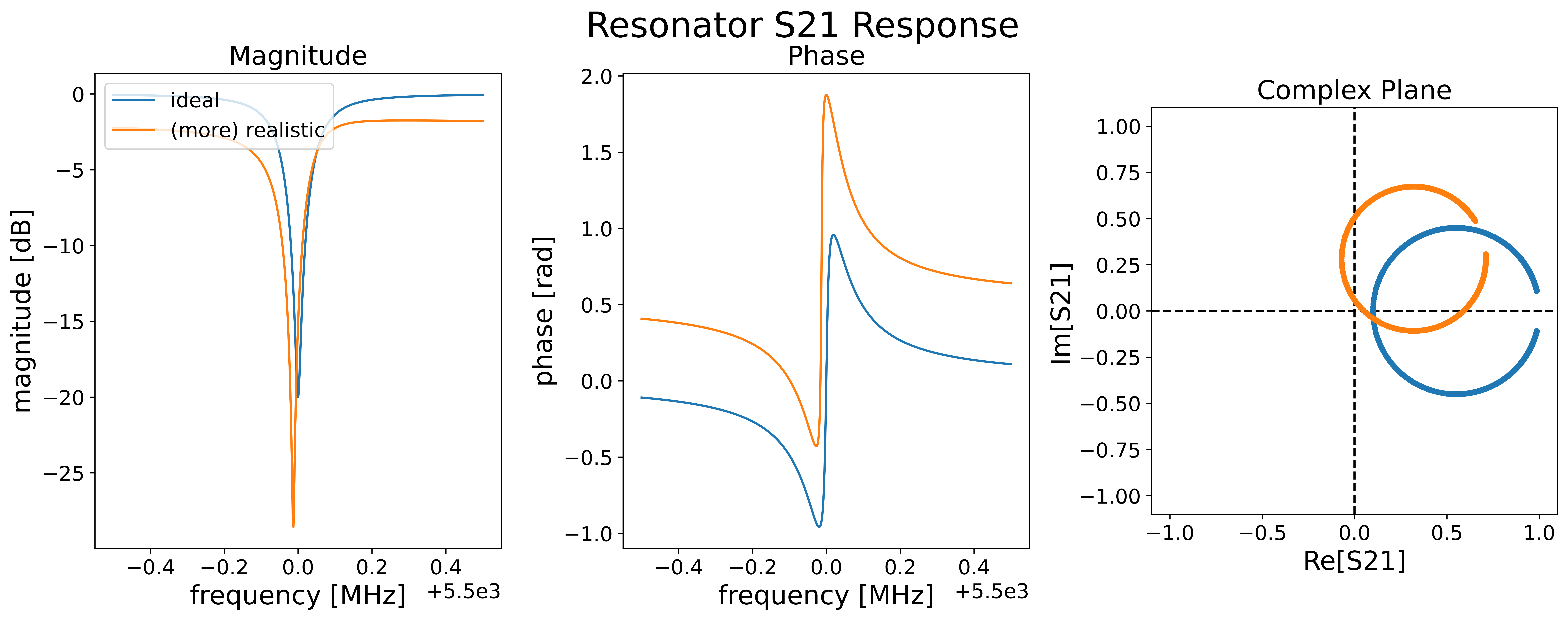}
    \caption{Amplitude (left), phase (center), and complex (right) $S_{21}$ responses for an ideal resonator (blue) and a more realistic resonance with some asymmmetry, loss, and phase delay (orange). Note that these effects rotate and scale the resonance circle in the complex plane away from the ideal response. Here we add a small asymmetry in the form of a complex coupled $Q$, 3 dB of loss from various connections, and 30$^\circ$ of delay corresponding to about 0.5mm of cable length drift. }
    \label{fig:s21ex}
\end{figure}

A resonator's complex transmission response as a function of frequency $S_{21}(f)$ may be parameterized by an amplitude and phase, where the resonance exhibits a dip in amplitude and a sign change in phase at the resonance frequency $f_\text{res}$. 
Equivalently, the complex response can be separated into in-phase $I$ and quadrature $Q$ components that are orthogonal to each other. 
An ideal resonator response in the $IQ$ plane traces out a circle oriented such that at the resonance frequency, the response is entirely in the quadrature direction. 
Non-idealities such as cable delays, impedance mismatches on the transmission line, and coupling to box modes may in general rotate or scale the response such that the response at the resonance frequency is no longer contained entirely in one axis. 
A sketch of the various parameterizations for an ideal and non-ideal resonator is given in Figure~\ref{fig:s21ex}. 

\section{SMuRF electronics}
\label{sec:smurf}

This section briefly outlines the relevant details of the SLAC Microresonator RF (SMuRF) electronics. 
The interested reader is directed to other SMuRF publications for further details.\cite{henderson18,yu22}

The SMuRF electronics system is a warm readout system designed for use with cryogenic resonator-based arrays. 
It generates the RF probe tones used to interrogate the resonances and reads back the transmitted RF response. 
The system additionally provides cryogenic amplifier and detector biases, flux ramp signal generation, timing synchronization, and data streaming to comprise a complete \umux readout system for large-scale experiments. 
The firmware on the main FPGA contains the bulk of the advanced signal processing, including RF tone generation and readback, signal manipulation, and clock synchronization. 
In the text that follows we describe the SMuRF firmware that was optimized for readout of CMB-style \umux resonators, which have bandwidths of $\sim 100$~kHz and resonance frequencies in the $4-8$~GHz range. 
Other applications may have slightly different bandwidths and channel counts, but the implementation details remain the same. 

The RF probe tone generation and readback is implemented in firmware as an oversampled-by-2 polyphase filter bank converting between 500~MHz-wide chunks of bandwidth digitized by each ADC and overlapping 2.4~MHz-wide channels, each of which may be mapped to a single detector. 
For each channel, SMuRF simultaneously reads out the complex response $(I,Q)$, where the orthogonal components separation is done digitally via Digital DownConversion (DDC) rather than relying on the orthogonality and stability of physical $I/Q$ signal chains. 

Since the detector signal is encoded in the resonance frequency of the corresponding resonator, following calibration as described in Subsec~\ref{subsec:freqest} the SMuRF firmware translates the $(I,Q)$ digitized RF response to a resonance frequency estimate. 
This frequency estimate is used to tune the probe tones to excite the resonators. 
The linearity requirements for the RF hardware, particularly the cryogenic amplifiers and RF mixers, in a system reading out hundreds or even thousands of resonators are quite strict. 
To address this, SMuRF system uniquely implements a tone-tracking feedback algorithm that continuously updates the probe tone frequency to maintain its centering on resonance. 
This minimizes tone power transmitted to the cryogenic amplifier, since the resonator response acts as a filter attenuating the probe tone. 

Finally, since in \umux systems the detector signal is encoded in the phase of the flux ramp-modulated resonance frequency, the SMuRF system demodulates the resonance frequency signal as described in Subsec.~\ref{subsec:frdemod}. 
It therefore outputs the demodulated detector signal, which can be streamed to experiment data acquisition programs or written to disk.

\subsection{Frequency Estimation}
\label{subsec:freqest}

As discussed in Subsec.~\ref{subsec:resreadout}, the complex response of an ideal resonator is centered in the $I/Q$ plane such that small deviations from the resonance frequency are contained in a single quadrature. 
However, non-idealities generically rotate and scale the response to an arbitrary position in the $I/Q$ plane. 
For each resonator, the complex response $S_{21}(f)$ is measured carefully, and a probe tone centered on the resonance frequency is applied. 
Currently, the probe tone is chosen to sit at the point of minimum $S_{21}$ amplitude response, though other choices are possible. 

The SMuRF estimates the deviation of the resonance frequency from the probe tone $\Delta f$ by applying a calibration factor to the measured ADC digital $(I,Q)$ response such that resonator frequency variation contributes only in one quadrature. 
In the small-signal limit where the frequency shift $\Delta f$ is small compared with the resonance bandwidth, we take the simplifying assumption that shifting the resonance frequency from the probe tone is the inverse of shifting the probe tone relative to the resonance at constant flux, i.e.

\begin{equation}
\label{eq:dfassumption}
	\frac{\partial S_{21}(f - f_\text{res}(\phi))}{\partial f}\bigg{|}_{\phi = \textrm{constant}} = - \frac{\partial S_{21}(f - f_\text{res}(\phi))}{\partial f_\text{res}(\phi)}.
\end{equation}

\noindent The analogous statement for KID statements modulates $S_{21}$ as a function of power on the detector rather than flux. 
As a consequence of this assumption, for reasonably constant conditions the resonator response $S_{21}$ must only be mapped once at a constant flux, recording the $I$ and $Q$ components of the transmission. 

The resonator response is rotated and scaled such that the response is in a single quadrature, allowing for a single real-valued estimate $\hat{\Delta f}$. 
The calibration factor $\eta$ is estimated as the inverse of the unit modulus transmission measured at some frequency offset $f_\pm \equiv f_\text{res} \pm f_\text{offset}$ from the resonance frequency. 
That is, $\eta$ is given by

\begin{equation}
	\label{eq:etaest}
	\eta\equiv\frac{f_+ - f_-}{S_{21}(f_+) - S_{21}(f_-)}
\end{equation}

\noindent The value of $f_\text{offset}$ is chosen empirically to capture the flat portion of the resonator circle; it is typically about 10\% of the resonance bandwidth. 

\begin{figure}
\label{fig:etause}
\includegraphics[width=0.95\textwidth]{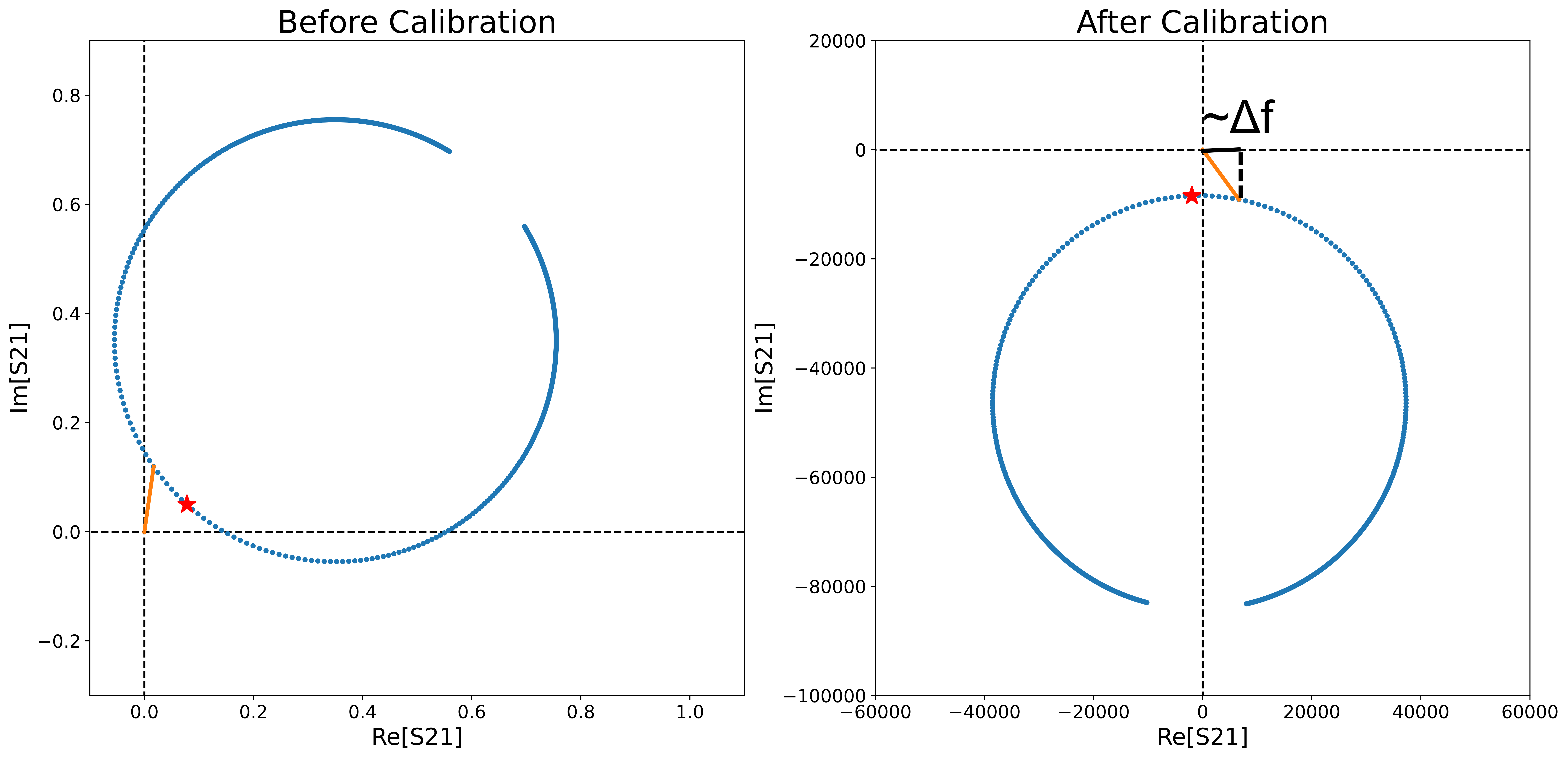}
\centering
	\caption{(Left) A resonance circle prior to calibration. (Right) A resonance circle after $\eta$ has been applied. The red stars denote the resonance frequency, while the orange line denotes the probe tone. Note that the axis scales have changed, since the $\eta$ calibration involves a magnitude scaling into frequency units. The difference between the probe tone and resonance frequency is estimated by the projection of the response onto a single axis. This difference $\Delta f$ is subsequently minimized in the tracking loop.} 
\end{figure}

Near resonance, we may then estimate the resonance frequency as 

\begin{equation}
	\label{eq:deltafest}
	\hat{\Delta f} \equiv \text{Re}\left[S_{21}(f_\text{probe} + \Delta f) \times \eta\right]
\end{equation}

\noindent where $S_{21}(f + \Delta f)$ is the complex transmission at the shifted resonance frequency. 
We note that the real or imaginary component taken in Eq.~\ref{eq:deltafest} depends on the convention used for the resonator $I/Q$ plane relative to the quadratures obtained from the ADC and DDC. 
Other documents may use a different convention, resulting in the imaginary component being taken rather than the real. 
A sketch of the $\eta$ estimation and application is given in Fig.~\ref{fig:etause}. 

With this frequency error estimate, the SMuRF feedback loop seeks to minimize $\hat{\Delta f}$ by continuously updating the probe tone frequency $f_\text{probe}$ such that $\Delta f\to 0$. 
This keeps the probe tone centered on the resonance dip as the resonator is modulated, thereby minimizing probe tone power transmitted to the cryogenic amplifier. 
The resonance frequency as a function of time $f_\text{res}(t)$ is thus estimated as the tracked (probe tone) frequency. 

We emphasize that the estimate relies on proper calibration of $\eta$, which in principle need be computed only once at the start of observation. 
However, any cable delay drifts or other time-varying changes to the resonator response induce rotations or scalings of the resonance circle not captured by $\eta$, potentially causing unlocked feedback or improper reconstruction of detector signals. 
This presents a tradeoff between more frequent recalibration and observing efficiency. 
To address this, slow changes to resonator response may be monitored with ``pilot tones'' monitoring the overall delay rather than individual resonators, and a slower feedback loop may be implemented to update $\eta$ without the need for recalibration.\cite{maxpilottones}

\subsection{Flux Ramp Demodulation}
\label{subsec:frdemod}

Following $\eta$ calibration, the feedback loop has a bandwidth of $\mathcal{O}$(1~kHz), allowing for tracking of slow modulations of the resonance frequency. 
This bandwidth may be sufficient for KID applications, but for \umux the flux ramp modulation is typically run at SQUID modulation rates of $\mathcal{O}$(10s~kHz) to adequately sample the science signal, suppress $1/f$ noise from resonator frequency fluctuations, and avoid aliasing penalties.\cite{ltd21}
To extend the bandwidth of the tracking loop to these frequencies and simultaneously demodulate the detector-equivalent signal from the flux ramp modulation without need for saving to disk the full modulated timestreams, SMuRF implements a stochastic gradient descent algorithm to minimize $\Delta f$ via a least mean squares fit to the harmonics of a known flux ramp modulation frequency. 

Given the quasi-sinusoidal nature of the SQUID flux to resonance frequency relation, the flux ramp-modulated resonance frequency may be approximated as a Fourier series with known frequency but unknown amplitude and phase. 
The principle harmonic $\omega_c \equiv 2\pi f_c$ is given by the flux ramp sawtooth reset rate multiplied by the number of a flux quanta swept in a ramp period, typically 3-6. 
Thus, the resonance frequency is parameterized by 

\begin{equation}
\label{eq:frvt}
f_\text{res}[n] = \text{constant} + \sum_{i=1}^M \left(a_i \sin\omega_i T_s n + b_i\cos\omega_i T_s n\right) + w[n]
\end{equation}

\noindent where $w[n]$ is a white noise term and $\omega_i$ are the $i$th harmonics of the principle harmonic $\omega_c$. 
Note that we write this in discrete time given the discretely sampled nature of the waveform as measured by the ADC; thus $n$ takes on only integer values. 
We may then estimate the phase of the $i$th harmonic and therefore the inferred TES current from the harmonic coefficients $a_i,b_i$ via $\Delta\arctan(b_i/a_i) = i\Delta \phi_i$. 
Currently SMuRF tracks and estimates up to $M=3$ harmonics, which are empirically sufficient to maintain good tracking and measurement of the SQUIDs tested thus far. 

We estimate the coefficients $a_i,b_i$ via a stochastic gradient descent method implemented in firmware. 
Equation~\ref{eq:frvt} is rewritten in vector form as

\begin{equation}
\label{eq:frvtmat}
\vec{f} = \mathbf{H}\vec{\alpha} + \vec{w}
\end{equation}

\noindent for $\vec{f}$ and $\vec{m}$ the resonance frequency and measurement noise, respectively of $M$ discrete samples in a flux ramp frame. 
Coefficient vector $\vec{\alpha}$ is a $(2M+1)\times 1$ vector of ($a_1,b_1,\ldots,a_M,b_M$, constant) coefficients for the sine and cosine coefficients of the $M$ harmonics and constant term. 
This is multiplied by the harmonic sample matrix $\mathbf{H}$, an $N\times (2M+1)$ matrix where the $(i,j)$ entries are given by

\begin{equation}
\label{eq:hmat}
H_{ij} = \left\{\begin{tabular}{cc}
$\cos_{(j+1)/2}[i],$&\quad $j$ = odd\\
$\sin_{j/2}[i],$&\quad $j$ = even\\
1,&\quad $j = 2M + 1$
\end{tabular}\right.
\end{equation}
\noindent for
\begin{equation}
\label{eq:sincomp}
\sin_m[n] \equiv \sin(\omega_m T_s n),\quad \omega_m = m\times \omega_1
\end{equation}
\noindent and likewise for $\cos_n[m]$. 
Here $T_s$ is the time per discrete sample. 
In the case of no flux ramp modulation, we have $N=0$ and thus $\vec{\alpha}$ reduces to a single constant $\alpha = \Delta f$. 

At a discrete timestep $n$, we then predict the resonance frequency from Eq.~\ref{eq:frvt} as 

\begin{equation}
\label{eq:festn}
\hat{f}[n] = \textrm{constant} + a_1\sin\omega_1 T_s n + b_1\cos\omega_1 T_s n + \cdots
\end{equation}

\noindent We see that this is equivalent to the dot product of the $n$th row of $\mathrm{H}$ with $\vec{\alpha}$, i.e.

\begin{equation}
\label{eq:festnmat}
\hat{f}[n] = \vec{H}_{n*}\cdot\alpha
\end{equation}

\noindent where the $*$ runs over all indices. 

We estimate the frequency error between the probe tone and the resonance frequency $\hat{\Delta f}[n]$ as in Subsec.~\ref{subsec:freqest} using the $\eta$ calibration. 
This error term allows us to update the prediction to minimize $\hat{\Delta f}$ via a stochastic gradient descent:

\begin{equation}
\label{eq:fb}
\vec{\alpha}[n+1] = \vec{\alpha}[n] + \mu \hat{\Delta f}[n]\left(\vec{H}_{n*}\right)
\end{equation}

\noindent where $\mu$ is a user-defined gain. 
Since $\omega_1$ is input by the user, this allows the tracking loop to have bandwidth beyond the constant-only case. 
Given the frequency error term accumulates the phase error estimate over the entire flux ramp frame, we may think of the feedback as somewhat analogous to a PID controller with only a proportional term. 

Given the harmonic coefficients, the phase of the first harmonic is computed as $\phi_1 = \arctan(b_1/a_1)$. 
This is typically averaged over the flux ramp frame and passed to a downstream data framer or external data acquisition system, which may further downsample or filter the data as needed. 
We thus consider the flux ramp-demodulated data output from SMuRF as sampled at the flux ramp sawtooth reset rate, though the harmonic coefficients are updated at the per-channel digitizer rate. 
The effects of these various sample rates are discussed further in Subsec.~\ref{subsec:alias}.

\section{Software Implementation}
\label{sec:software}

The frequency estimation, tracking, and flux ramp demodulation components of the SMuRF firmware algorithm are critical to transforming the measured resonator response to the inferred detector-equivalent signal. 
To enable exploring the impact of these algorithms and sensitivity to user-defined parameters, we have implemented a software simulation suite, \texttt{babysmurf}\footnote{\texttt{https://github.com/cyndiayu/babysmurf}}, which returns outputs given known inputs. 

We note that these components act on a per-channel basis.
Thus, any effects of the channelization or considerations of interacting channels such as crosstalk or colliding resonance  modulations are not currently modeled. 
Since the channelization filter bank is the inverse of the mechanism by which probe tones are generated, this implies that we ignore the finite width of the probe tone and here consider it as a spectrally pure tone with infinite dynamic range. 
We furthermore do not implement any SQUID or resonator physics; arbitrary shapes may be generated without enforcing that they are physically plausible and we do not account for any changes to the SQUID or resonator response as a function of input power. 
These features are planned for future updates and are welcomed as contributions from other users. 

\begin{figure}
\centering
\label{fig:flowchart}
\includegraphics[width=0.95\linewidth]{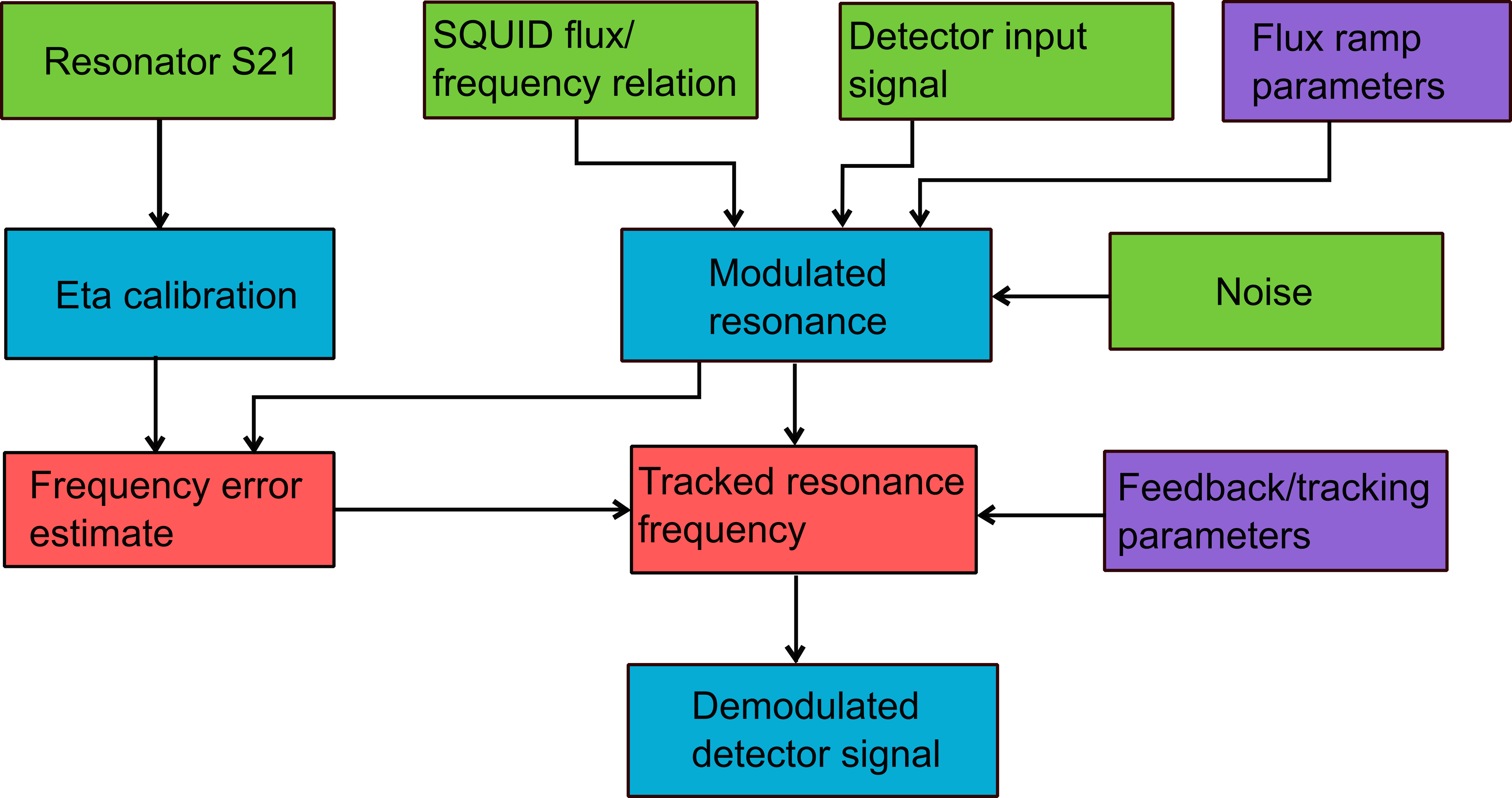}
\caption{A flowchart of the \texttt{babysmurf} software suite. The green components are physical inputs, which may come from real data or be generated given key parameters. The purple components are flux ramp, feedback, and tracking parameters that must be input by the user. From these, the blue components are calculated. The red components represent the main non-trivial algorithms implemented by SMuRF.}
\end{figure}

A flowchart of the software is given in Figure~\ref{fig:flowchart}. 
Resonator response, SQUID parameters, detector signals, and noise may be input by the user from real data, or generated based on defining parameters. 
The flux ramp and feedback/tracking variables are implementation-specific parameters required to be specified by the user. 
From these inputs, the calibration factor and flux-modulated resonance frequency response may be calculated. 
The core SMuRF algorithms compute the frequency error estimate and tracked resonance frequency, from which the demodulated detector signal is derived. 

\subsection{Generating Resonances}
We model the forward transmission of a resonance with total quality factor $Q$ and coupled quality factor $Q_c$ as

\begin{equation}
	\label{eq:s21eq}
	S_{21}(f) = 1 - \frac{Q*Q_c^{-1}}{1 + 2iQ(f-f_\text{res})/f_\text{res}}
\end{equation}

\noindent where $f_\text{res}$ is the resonance frequency.\cite{khalil12}
The coupling quality factor term $Q_c$ may be complex, which allows for asymmetric resonances due to impedance mismatches between the resonator input and output ports. 
While the resonator parameters being considered for various applications may vary widely, we take as default parameters the behavior of NIST CMB \umux resonators, which have typical quality factors $Q\sim 5\times 10^4$ and resonance frequencies around $f_\text{res}\sim 5~\text{GHz}$.\cite{dober21}
Throughout the code, we parameterize the complex resonator response via a separate magnitude and phase to avoid confusion of conventions on the quadrature labels. 

For a generated resonance, we may estimate $\eta$ as described in Eq.~\ref{eq:etaest}. 
This allows exploration of the accuracy of the frequency error estimation as a function of parameters such as resonance asymmetry and offset frequency, which may in turn inform user choices of estimation parameters given a particular realization of resonators. 

\subsection{Flux Ramp Modulation}
The flux ramp-modulated resonance frequency as a function of time is parametrized by a carrier frequency $\omega_c$ as

\begin{equation}
	\label{eq:sqcurve}
	\Delta f_\text{res}(t) = C\left(\frac{\lambda\cos(\omega_c t)}{1 + \lambda\cos(\omega_c t)}\right)
\end{equation}

\noindent where $\lambda$ is the SQUID hysteresis parameter. 
We note that this equation is not fully physical, and the full modeling of SQUID curves is the subject of intense study.\cite{wegner22}
For the purposes of the simulation suite, however, it is sufficient to capture the quasi-sinusoidal nature of SQUID curves often encountered. 
The value of $C$ may be scaled such that the overall peak to peak swing of the resonance frequency $\Delta f_\text{pp}$ is appropriately matched to the bandwidth. 
We typically set $\Delta f_\text{pp}\sim 100~\mathrm{kHz}$. 
By appropriately scaling $t$ we may reparameterize Eq.~\ref{eq:sqcurve} in terms of input flux, allowing for arbitrarily modifying the sample rate or number of $\Phi_0$ swept in a flux ramp frame. 
A SQUID curve in units of resonance frequency versus input flux can similarly be measured from a cold system and passed as input to the simulation suite. 

Given a SQUID curve covering a full $2\pi$ in input flux, a flux ramp modulated signal may be generated by perturbing the phase of each flux ramp frame, where one frame is a single sawtooth ramp. 
The phase modulation is achieved by shifting the flux ramp frame forward and backward by an integer number of samples. 
Thus, the phase modulation is eventually discretized by the finite sampling of the SQUID curve, which should be kept in mind when interacting with the software. 

Given a detector-equivalent signal as a function of time, it is interpolated or downsampled such that it is sampled at the flux ramp frame rate. 
This does not enforce the continuity of the resonance frequency at sawtooth resets, which may have large transient signals or other effects of large input signals. 

The modulation of the resonance $S_{21}$ is implemented na\"ively as a modulation of the entire resonance shape in frequency space, with no incorporation of SQUID or resonator dynamics. 
That is, the entire resonator $S_{21}$ is modulated forward and backward in frequency without changing its shape in amplitude or phase.

\subsection{Demodulation and Output Inspection}
The tracking and demodulation algorithm as described in Subsec.~\ref{subsec:frdemod} is implemented in two modes: (1) taking only the flux ramp-modulated resonance frequency as input and assuming perfect tracking, and (2) taking the $\eta$ estimate and modulating $S_{21}$ as input such that the frequency error estimation is also used. 
This allows for exploration of the effects of the gradient descent and demodulation algorithms separately from the effects of the frequency error estimation if desired. 
The fitted coefficients $\vec{\alpha}[n]$, tracked frequencies, estimated and real frequency errors, and demodulated phase are all recorded for later inspection and comparison.

\subsection{Noise}
Noise is added at the timestream level by drawing from a distribution according to a user-defined noise amplitude spectral density expressed in units of frequency noise (Hz/$\sqrt{\text{Hz}}$). 
The phases are randomly drawn from a uniform distribution $[0,2\pi)$. 
An empirically measured noise trace may also be added in manually. 
This allows the user to disentangle the effects of the tracking algorithms from noise, and conversely to investigate the sensitivity of the tracking and estimation loops to noisy inputs. 

\subsection{Jupyter Notebooks}
An introductory notebook outlining the basic functionality of the software and example notebooks reproducing the results of Sec.~\ref{sec:ex} are available alongside the main software module. 
The main dependency is \texttt{numpy}.\cite{numpy}
Additional features are enabled by \texttt{scipy} and \texttt{matplotlib}.\cite{scipy,matplotlib}

\section{Ongoing Example Studies}
\label{sec:ex}
We present several short studies that have been enabled by this simulation suite. 
Further directions for exploration are discussed in Sec.~\ref{sec:conclusion}. 

\subsection{Eta miscalibration}
\label{subsec:etaoff}
As the $\eta$ calibration discussed in Subsec.~\ref{subsec:freqest} determines the frequency error estimate that is minimized in tone tracking, the accuracy and stability of this calibration factor is key to the readout scheme. 
There are several ways in which the calibration factor may be wrong: it could be misestimated, as is often the case particularly for very shallow resonances where the flat portion of the resonance circle is very small; the resonator response may be drifting due to time-varying loss or phase delay without an updated calibration factor; it may be estimated using an algorithm which fails to capture the point of maximum responsivity, as may be the case for asymmetric resonances. 
These various effects are not guaranteed to have the same effect on the flux ramp-demodulated data, and must be considered separately. 

We consider here the impact of a phase delayed resonance whose calibration factor has not been updated. 
That is, the resonance circle has been rotated, perhaps due to temperature-induced cable length variation or other slowly-varying effect, without a compensating update in the calibration parameter. 
Here we take the resonance to be perfectly symmetric and noiseless such that the only non-ideality is the rotation of the resonance circle relative to its initial orientation. 
The rotation is implemented by adding a constant phase delay to the entire modulated $S_{21}$ response; thus, this does not capture time-varying delays. 

\begin{figure}
\centering
\label{fig:eta_offset}
\includegraphics[width=0.95\linewidth]{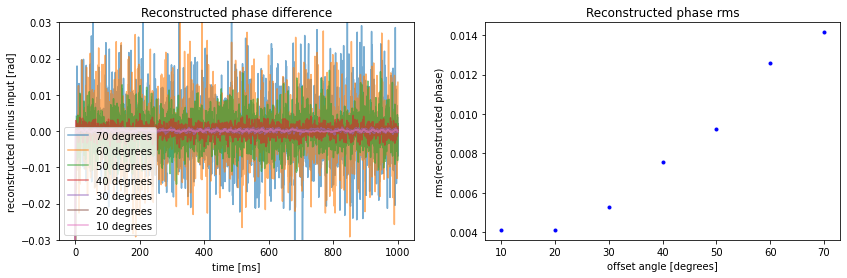}
\caption{(Left) Differences in flux ramp-demodulated timestreams for scenarios in which the $S_{21}$ response has been shifted by some constant phase delay, but the $\eta$ calibration has not been updated to reflect this delay. The reconstructed timestreams have an effective noise bias that grows with the size of the angle offset. (Right) The rms error of the demodulated timestreams versus offset angle. We see that the amplitude of the noise is non-linear in the angle of offset.}
\end{figure}

We see that in this limit of a perfectly symmetric resonance and noiseless modulation, the impact of a misestimated calibration parameter appears as an effective noise. 
Note that since the feedback algorithm contains no randomness, this response is deterministic up to the choice of starting point for the $\vec{\alpha}$ coefficients. 
The pseudonoise has mean zero, suggesting that this effect may produce a noise penalty but no bias on the final science data. 
It is worth exploring in future studies how these results change for a time-varying delay, as may be the case for temperature-induced cable changes, or for asymmetric resonances which may see some bias in addition to the noise penalty. 

\subsection{Gain}
\label{subsec:gain}
\begin{figure}
\centering
\label{fig:example_gain}
\includegraphics[width=0.95\linewidth]{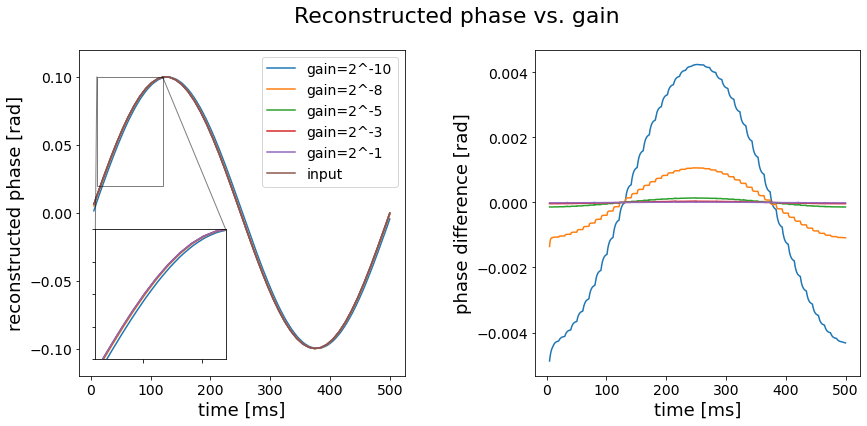}
\caption{(Left) Flux ramp-demodulated timestreams for an input signal under the assumption of perfect tracking and an ideal resonator, varying the gain. We see that the differences from the input signal are small across nearly 10 bits of precision in the gain setting. (Right) The error between the tracking algorithm reconstruction of the signal and the input. We see that the errors are sub-percent across a broad range of gains. The cusped nature of the error comes from the discrete sampling of the SQUID curves in the fake signal construction and is not expected to appear in real data. }
\end{figure}

The gain parameter determines the extent to which the tracking algorithm responds to the estimated frequency error between the probe tone and resonance frequency. 
If it is too small, the phase estimate may never converge over the course of a flux ramp cycle, while too large estimates render the tracking loop unstable. 
In Figure~\ref{fig:example_gain} we inspect the response to an input sine wave for values of gain spanning several orders of magnitude. 

We see that the tracking loop is approximately equivalent for gain values spanning several orders of magnitude. 
The small sawtooth-like features are a numerical artifact due to the finite sampling resolution of the input SQUID curve, and are not expected to appear in real data.

\subsection{Aliasing}
\label{subsec:alias}

The transient induced by the sawtooth reset on the flux ramp line breaks the quasi-sinusoidal assumption of the modulated resonance frequency and may cause the feedback loop to become unlocked. 
To avoid these effects, a user-defined portion of each flux ramp frame is blanked off from the tracking loop: the resonance frequency is held constant and the $\alpha$ coefficients are not updated. 
In typical operation, we attempt to tune the flux ramp waveform amplitude and blank-off such that the region covered by the tracking loop sweeps an integer number of flux quanta and thus the quasi-sinusoidal assumption holds. 

\begin{figure}
\centering
\label{fig:aliasing}
\includegraphics[width=0.95\linewidth]{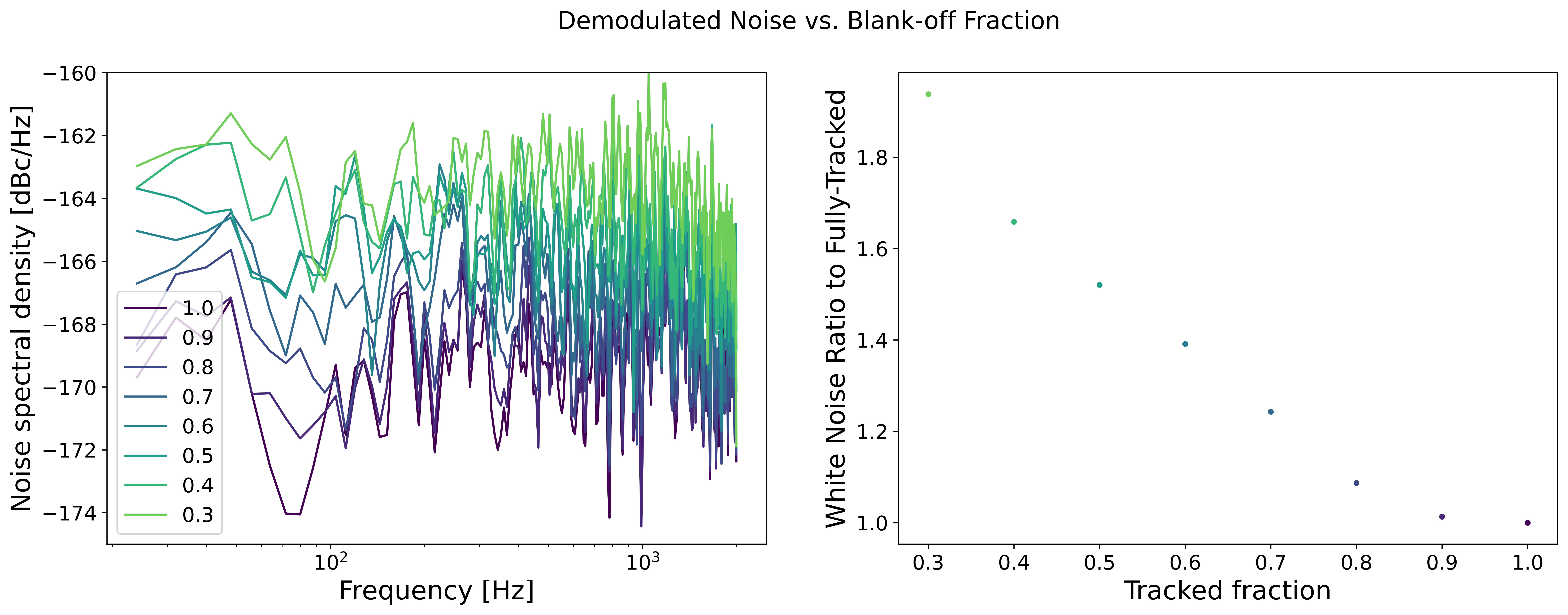}
\caption{(Left) Noise spectral densities for the same noisy timestream passed to the demodulation loop, but with varying fractions of each flux ramp frame tracked. Blanking off the flux ramp frame allows for transients due to the flux ramp sawtooth reset frames to be ignored by the tracking loop, maintaining a quasi-sinusoidal response. (Right) The ratio of the white noise levels (in amplitude spectral density units of Hz/$\sqrt{\mathrm{Hz}}$) to the case of perfect tracking the full waveform. }
\end{figure}

This blank-off induces an aliasing penalty during the region not covered by the tracking loop, which manifests as an increase in noise. 
In Figure~\ref{fig:aliasing} we generate a white noise timestream and pass it through demodulation with a simulated SQUID curve. 
The absolute level of the noise does not strongly impact the simulated behavior so long as the equivalent flux noise is small compared to a $\Phi_0$. 
We vary the fraction of the SQUID curve that is tracked and demodulated and compare to the case of no aliasing, when the entire SQUID curve is used. 
Here, we assume perfect tracking and a symmetric resonator. 

We see that the aliasing penalty is sharper than the na\"ive $\sqrt{1/\text{tracked fraction}}$ expectation, possibly due to the discontinuities that arise from the fractions here not being chosen to maintain an integer number of $\Phi_0$ covered by the tracking loop.\cite{mates12}
Future studies may examine the impact of this blanking fraction choice as well as how it may depend on the shape of the SQUID curve. 

\section{Conclusion and Future Directions}
\label{sec:conclusion}
We hope and intend for this simulation suite to be a starting point for a wide array of investigations into the effects of the SMuRF readout electronics. 
The advanced RF hardware, fast tone updating, high dynamic range, and large-scale integration support of SMuRF make it an attractive solution for microwave resonator-based readout applications across a wide range of science goals. 
The SMuRF system's unique tone-tracking ability further enables high channel count applications and dramatically simplifies offline analysis for frequency estimation.
In the case of \umux, tone tracking additionally simplifies flux ramp demodulation. 

However, the effects of these algorithms may have subtle impacts on the inferred detector data that may include a noise penalty or systematic biases. 
The results of these simulations may be used to constrain design specifications for hardware, prepare systematic mitigation analyses, or design new algorithms as the applications dictate. 
Below we suggest several questions that may be answered by the existing software suite:
\begin{itemize}
\item Is the tracking linear for all inputs? For what gain settings does this linearity hold? More generally, under what conditions is the demodulated output an unbiased estimate of the detector input signal? 
\item Are there better estimators for the resonance frequency or frequency error, particularly for asymmetric resonators?
\item Does the non-linearity of the flux ramp waveform due to DAC effects bias the demodulation? 
\item How do the assumptions for frequency error estimation and tone-tracking need to change in the case of KID-like devices, where the size of the resonance circle may be changing with input power? 
\item Can the signal-to-noise ratio be improved by tracking more harmonics? 
\end{itemize}

Furthermore, extensions to the software suite are welcomed and would enable analyses of other aspects of the readout, with an ultimate goal of providing an end-to-end simulation of the effects of the warm readout electronics. 
Avenues for further development include:
\begin{itemize}
\item Incorporation of resonator and SQUID physics effects, such that the modulation of tone power on the resonator input can be modeled. 
\item Integration with existing \texttt{pysmurf}\footnote{\texttt{https://github.com/slaclab/pysmurf}} tools for data-taking and analysis. 
\item Multi-channel effects, including but not limited to: colliding resonators and crosstalk. 
\item Channelization effects and the polyphase filter bank, such that the finite spectral width of the probe tones can be modeled with the resonator transfer function. 
\end{itemize}

The software repository will be maintained by the authors, with contributions highly welcomed. 
Questions about use and startup may be directed to the corresponding author. 

\acknowledgments 
CY was supported in part by the National Science Foundation Graduate Research Fellowship Program under Grant No. 1656518. 
MSF was supported in part by the Department of Energy Office of Science Graduate Student Research (SCGSR) Program. The SCGSR program is administered by the Oak Ridge Institute for Science and Education (ORISE) for the DOE, which is managed by ORAU under contract number DE-SC0014664. 
This work was supported in part by the Department of Energy at SLAC National Accelerator Laboratory under contract DE-AC02-76SF00515. 

\bibliography{spie} 
\bibliographystyle{spiebib} 

\end{document}